\documentclass[aps,prd,twocolumn,nofootinbib,floatfix]{revtex4-1}
\usepackage{setspace}
\usepackage{amsthm}
\theoremstyle{definition}
\newtheorem*{definition}{Definition}
\usepackage{graphicx}
\usepackage{dcolumn}
\usepackage{multirow}
\usepackage{subfigure}
\usepackage{times,mathptm}
\usepackage{float}
\usepackage{color}
\usepackage{amsmath,amsfonts}
\usepackage{mathptmx}
\usepackage{mathrsfs}
\usepackage{bbm}
\usepackage{bm}
\usepackage{xfrac}
\newtheorem{theorem}{Theorem}
\theoremstyle{definition}
\newcommand{\beq}{\begin{equation}}
\newcommand{\eeq}{\end{equation}} 
\newcommand{\bea}{\begin{eqnarray}}
\newcommand{\eea}{\end{eqnarray}} 
\newcommand{\Sc}{S$_\text{c}$}
\newcommand{\bx}{\mathbf{x}}
\newcommand{\by}{\mathbf{y}}

\newcommand{\tr}{\text{Tr}}
\renewcommand{\b}{\beta}

\newcommand{\q}{\overline{q}}

\newcommand{\vx}{\bx}
\newcommand{\vy}{\by}
\newcommand{\vz}{\vec{z}}

\newcommand{\Hs}{H_\text{spin}}
\newcommand{\Zs}{Z_\text{spin}}
\newcommand{\n}{\nu}
\newcommand{\m}{\mu}
\newcommand{\g}{\gamma}

\newcommand{\N}{{\cal N}}

\newcommand{\oh}{\frac{1}{2}}

\newcommand{\dg}{\dagger}
\newcommand{\non}{\nonumber}

\newcommand{\rf}[1]{(\ref{#1})}
\newcommand{\ra}{\rightarrow}
\newcommand{\pa}{\partial}

\begin{document}

\title{Symmetry, Confinement, and the Higgs Phase}

\bigskip
\bigskip

\author{Jeff Greensite and Kazue Matsuyama}
\affiliation{Physics and Astronomy Department \\ San Francisco State
University  \\ San Francisco, CA~94132, USA}
\bigskip
\date{\today}
\vspace{60pt}
\begin{abstract}
We show that the Higgs and confinement phases of a gauge Higgs theory, with the Higgs field in the
fundamental representation of the gauge group, are distinguished both by a broken or unbroken realization of the global center subgroup
of the gauge group, and by the type of confinement in each phase.  This is color confinement in the Higgs phase, and a
stronger property, which we call ``separation-of-charge" confinement, in the confining phase.
\end{abstract}

\maketitle
\singlespacing

\section{Introduction}

   In this article we would like to address two very old questions in gauge field theory, for which we will propose new answers.  
  The first question is:  What is meant by a
``spontaneously broken gauge theory," in view of Elitzur's theorem \cite{Elitzur:1975im}, the work of Osterwalder and Seiler \cite{Osterwalder:1977pc}, Fradkin and Shenker \cite{Fradkin:1978dv}, Banks and Rabinovici \cite{Banks:1979fi}, and
the fact that all physical particles in, e.g., an SU(2) gauge Higgs theory, are color singlets \cite{Frohlich:1981yi,tHooft:1979yoe}?  The second question is: What is meant by the
word ``confinement"  in a theory (such as QCD) with matter in the fundamental representation of the gauge group?  In such theories the order parameters associated with pure gauge theories, i.e.\ non-vanishing string tension, `t Hooft loops, Polyakov loops,  and vortex free energies have apparently non-confining behavior.

\begin{figure}[htb]
 \centerline{\includegraphics[scale=0.6]{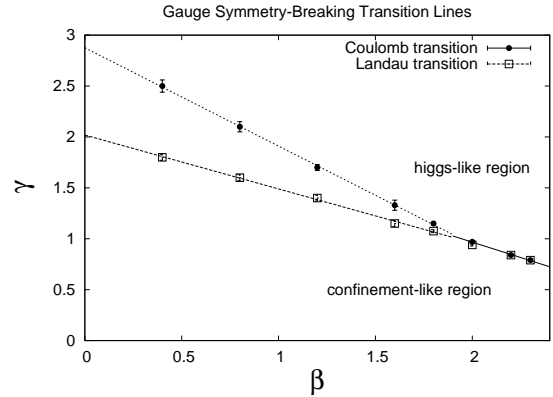} }
 \caption{The location of remnant global gauge symmetry breaking in Landau and Coulomb gauges, in the $\beta-\gamma$ coupling plane,
 for the SU(2) gauge Higgs theory in \rf{Sgh}. Figure from ref.\ \cite{Caudy:2007sf}.}
 \label{Billy}
\end{figure}

   Starting with the first question, most discussions of the Higgs mechanism begin with a ``Mexican hat" potential of some kind,  and the Higgs field is expanded around one of the minima $\phi_0$ of this potential, i.e.\ $\phi(x) = \phi_0 + \delta\phi(x)$,
where the particular minimum is selected by fixing to a unitary gauge.  When this replacement for $\phi(x)$ is inserted back into
the action,  some or all of the gauge vector bosons acquire a mass.  But then we may ask: in what sense is this spontaneous symmetry breaking?  In unitary gauge, for the U(1) and SU(2) groups, there is no symmetry left to break.  Perhaps, at least in connection with this issue, we should avoid gauge fixing? But in the absence of gauge fixing, a local gauge
symmetry cannot break spontaneously, as we know from Elitzur's famous theorem.  What about a middle way, i.e. choose some gauge, e.g.\ Coulomb, Landau, or axial gauge, which leaves unfixed a global subgroup (a ``remnant" symmetry) of the gauge group? A global symmetry \emph{can} break spontaneously.
This is perfectly consistent with Elitzur's theorem, and some textbooks and review articles do define ``spontaneous gauge symmetry
breaking" in this way.  The problem with that idea is that the location of the transition line is gauge dependent \cite{Caudy:2007sf}.  This is seen in Figure \ref{Billy}, where transition lines for the breaking of the global remnant symmetry were computed in SU(2) gauge Higgs theory in both Coulomb and Landau gauges.  
 
    A second reason to doubt that there is any essential distinction between the Higgs and confinement phases is the theorem proven by
Osterwalder and Seiler \cite{Osterwalder:1977pc} (see also the closely related work of Banks and Rabinovici \cite{Banks:1979fi}), whose consequences were elaborated by Fradkin and Shenker \cite{Fradkin:1978dv}.  This theorem states that in a lattice gauge Higgs theory with the Higgs in the fundamental representation, there is no thermodynamic phase transition which isolates the Higgs from the
confinement-like region;  one can always follow a path from a point in the one region of the phase diagram to the other without encountering a thermodynamic singularity.  And a third good reason, pointed out by Fr\"olich, Morcio and Strocchi \cite{Frohlich:1981yi},  `t Hooft \cite{tHooft:1979yoe},  and Susskind (cf.\ \cite{Banks:1979fi}), is that physical particles in the so-called ``Higgs" phase are created by local color singlet operators acting on the vacuum, as in the confinement region.  We will refer to 
a gauge theory in which all asymptotic particle states are color singlet objects as color (or ``C") confinement.  According to this definition both
the Higgs and confining regions of an SU(2) gauge Higgs theory, and also QCD, are C confining theories.

     Then is there any meaning to the word ``confinement" beyond C confinement?  In a pure gauge theory, of course there is.  There is the
area law for Wilson loops, the vanishing of the Polyakov line order parameter, the formation of stable color electric flux tubes between static sources,
and the consequent linear rise of the static quark potential with quark-antiquark separation.  In such theories, the confinement phase can be identified as the phase of unbroken center symmetry (distinct from the gauge symmetry).  However, in a gauge theory with matter in the fundamental representation,
none of these properties hold, at least not exactly or asymptotically.  There is no global center symmetry distinct from the gauge symmetry, long flux tubes are unstable due to string-breaking by matter fields, and the linearly rising potential eventually goes flat.

    In view of all these facts, we believe the informed consensus on the two questions we have raised is as follows:
 \begin{enumerate}
\item  Confinement, in a gauge + matter theory, can only mean that all of the
particles in the asymptotic spectrum are color neutral; i.e. confinement is C confinement. 
Both the confining and the
Higgs regions of a gauge Higgs theory are C confining. \vspace{5pt}
\item There is no such thing as a  ``spontaneously broken gauge symmetry," at least none that has any physical meaning.
In a gauge Higgs theory, the  ``Higgs" and ``confinement" regions of the
phase diagram are part of the same Higgs-confinement phase.
\end{enumerate}
In our opinion, these consensus views are both wrong.  

In the next section, we point out that the Higgs phase of a gauge Higgs theory is closely analogous to a spin glass, that the
relevant symmetry which is spontaneously broken is a global symmetry which transforms the Higgs field but does not transform the gauge field,
and this symmetry contains at a minimum the global center subgroup of the gauge group. This symmetry is distinct
from the quite different center symmetry whose order parameter is the Polyakov line.  We will construct a gauge invariant order
parameter, modeled after the Edwards-Anderson order parameter \cite{Edward_Anderson} for a spin glass, which can detect the spontaneous breaking
of this global subgroup of the gauge symmetry.  This order parameter does not depend, even implicitly, on a gauge choice.
In the following section 3 we introduce the concept of separation-of-charge (\Sc) confinement, which is a stronger 
condition than color confinement, and which generalizes the concept of confinement in pure gauge theories to theories with dynamical matter and string-breaking.  We will explain how, in the absence of a massless phase, the phase of
unbroken global center gauge symmetry is the \Sc\ confining phase, while the spin glass (aka Higgs) phase is a C confining phase, in
which this global symmetry is spontaneously broken. 

   This article is a review, and we concentrate on simply describing our main results.  The interested reader will find detailed derivations in the cited references.

\section{The Higgs phase as a spin glass}
  
    The Edwards-Anderson model of a spin glass is an Ising model with random couplings among spins, i.e.
\beq
           \Hs =  - \sum_{ij} J_{ij} s_i s_j - h \sum_i s_i \ ,
\eeq
where $s_i = \pm 1$, the sum is (usually) over nearest neighbors, and the $J_{ij}$ are a set of random couplings
between sites $i,j$ with probability distributions $P(J_{ij})$.  There is obviously a global $Z_2$ symmetry $s_i \ra \pm s_i$ in the $h\ra 0$ limit.  But after the sum over random couplings we have $\langle s_i \rangle \ra 0$ in this limit.  Despite that fact, there is still a way to detect the
spontaneous breaking of the global symmetry.  Following \cite{Edward_Anderson}, let us define
\bea
           \Zs(J) &=&  \sum_{\{s\}} e^{-\Hs/kT}  \label{sg1} \non \\
           \overline{s}_i(J) &=& {1\over \Zs(J)}  \sum_{\{s\}} s_i e^{-\Hs/kT} \non \\ 
           q(J) &=& {1\over V} \sum_i  \overline{s}^2_i(J)   \non \\
           \langle q \rangle &=& \int \prod_{ij} dJ_{ij} \ q(J) P(J) \label{sg4} \ ,
\label{q}
\eea  
 where  $q(J)$ is called the Edwards-Anderson order parameter. Its expectation value $\langle q \rangle$ is non-zero in the spin glass phase, indicating spontaneous symmetry breaking of the global $Z_2$ symmetry, despite the fact that $\langle s_i \rangle = 0$.

\begin{table}[H]
\begin{center}
\begin{tabular}{|c|c|c|c|} \hline
      Model                      & ``Spin" &  random coupling  &  global symmetry  \\ \hline
       Edwards-Anderson & $s_i$   &          $J_{ij}$            &      $Z_2$                \\  \hline
       gauge Higgs           &  $\phi(\vx)$    &   $U_k(\vx)$    &     custodial       \\ \hline
\end{tabular}
\end{center}
\caption{Analogies between spin glass and gauge Higgs theories.}
\label{analog}
\end{table} 

    The gauge Higgs analogy to a spin glass is shown in Table \ref{analog}.  
In a gauge Higgs theory the Higgs field $\phi$ plays the role of spin, and the gauge link variables $U_i$ play the role of random couplings.  
As in a spin glass, $\langle \phi \rangle=0$, and for similar reasons.  But this is not the end of the story, as regards spontaneous
symmetry breaking.

   We will define a {\it custodial} symmetry to be a group whose elements transform the Higgs field, but do not transform the gauge field.  It is
clear that this group contains, at a minimum, the global center subgroup of the gauge group.  Let us begin from the continuum action of a gauge Higgs theory
\beq
S = \int d^4x \bigg\{ {1\over 4} \tr F_{\mu \nu} F^{\m \nu} +  \oh \phi^\dg (-D^2) \phi + \lambda (\phi^\dg \phi - \g)^2 \bigg\} \ .
\eeq
On the lattice, for the SU(2) group, and taking the $\lambda \ra \infty$ limit we may write
\bea
     S   &=&  - \beta \sum_{plaq} \oh \mbox{Tr}[U_\m(x)U_\n(x+\hat{\m})U_\m^\dg(x+\hat{\n}) U^\dg_\n(x)]  \non \\
           & & \qquad - \gamma \sum_{x,\m} \oh \mbox{Tr}[\phi^\dg(x) U_\m(x) \phi(x+\widehat{\m})] \ ,
\label{Sgh}
\eea
 where $\phi$ is an SU(2) group-valued field.  The possibility of expressing $\phi$ as a group-valued field is special
 to SU(2), and the custodial symmetry transformations are
 \beq
 \phi(x) \ra \phi(x) R ~ , ~~~\mbox{where}~~~R \in SU(2) \ .
 \eeq
 This global SU(2) custodial group of course contains  $Z_2$ as a subgroup, which is indistinguishable from the 
 global $Z_2$ subgroup of the gauge group.  For larger SU(N) gauge groups, the custodial group may contain \emph{only} the
 global $Z_N$ center subgroup of the gauge group.
 
    Now to make the correspondence with the Edwards-Anderson treatment, let us first define
\bea
           \lefteqn{\exp[-H(\phi,U)/kT]} \non \\
            & & \qquad =  \langle \phi,U| e^{-H/kT} |\phi,U\rangle  \non \\
           & & \qquad =   \sum_n |\Psi_n(\phi,U)|^2 e^{-E_n/kT}  \non \\ 
           & & \qquad =  \int DU_0 [DU_i D\phi]_{t\ne 0}  \exp[-S(\phi(\vx,t),U_\m(\vx,t))]  \,
\eea
and
\bea
        H_{spin}(\phi,U,\eta) &=& H(\phi,U) -  h \sum_{\vx} \tr[\eta^\dg(\vx) \phi(\vx)]  \ ,
\eea
with $\eta(\vx)$ an SU(2)-valued field.  The term proportional to $h$ is introduced for formal reasons, since the proper
definition of spontaneous symmetry breaking requires first introducing a small explicit breaking term, then taking the infinite
volume limit, followed by $h\ra 0$.   We then define
\bea                        
            \Zs(U) &=& \int D\phi(\vx) \ e^{-\Hs(\phi,U,\eta)/kT}   \label{gg1}  \non \\                     
           \overline{\phi}(\vx;U) &=& {1 \over  \Zs(U) }  \int D\phi \ \phi(\vx)  e^{-\Hs(\phi,U,\eta)/kT}  \label{overline} \non \\ 
            \Phi(U) &=& {1\over V} \left[\sum_\vx  | \overline{\phi}(\vx;U) | \right]_{\eta \in \N(U)} \label{max}  \non \\
            \langle \Phi \rangle &=& \int DU_i(\vx) \ \Phi(U) P(U) \ ,
\label{Phi}
\eea
 where  
 \beq
     \N(U) = \underset{\eta}{\arg\max} \sum_\vx \left| \int D\phi \ \phi(\vx)  e^{-\Hs(\phi,U,\eta)/kT} \right| \ .
\eeq
Eq.\  \rf{Phi} for gauge Higgs theory should be compared with the Edwards-Anderson order parameter in \rf{q}.
The term proportional to $h$ is introduced in such a way (involving $\N(U)$) that it does not break local gauge invariance, cf.\ \cite{Greensite:2020nhg}.

We still have to specify $P(U)$, but there is only one possibility.  We must have, for the thermal
average of an operator $Q(U)$ on a time slice,  
\bea
            \langle Q \rangle &=&   {\tr \ Q e^{-\Hs/kT} \over \tr \ e^{-\Hs/kT}} = \int DU_i(\vx) \ Q(U) P(U) \ .
\eea
Since $\langle Q \rangle$ is the standard thermal average of $Q(U)$ on a time slice in a gauge Higgs theory, it requires that \bigskip
\beq
             P(U) = {\Zs(U) \over Z} \ .
\eeq
Both $\Zs(U)$ and $P(U)$ are gauge invariant,
even at finite $h$, and we are mainly interested in the zero temperature $T\ra 0$ limit.

We now have a gauge invariant criterion for the spontaneous breaking of custodial symmetry:
\bea
           \lim_{h \ra 0} \lim_{V \ra 0} \langle \Phi \rangle \left\{ \begin{array}{cl}
                           = 0 & \text{unbroken symmetry} \cr \cr
                          > 0  & \text{broken symmetry} \end{array} \right. \ ,
\eea
which is entirely analogous to the Edwards-Anderson criterion for the spontaneous symmetry breaking of
global $Z_2$ symmetry in a spin glass: 
\bea
           \lim_{h \ra 0} \lim_{V \ra 0} \langle q \rangle \left\{ \begin{array}{cl}
                           = 0 & \text{non-spin glass phase} \cr \cr
                          > 0  & \text{spin glass phase} \end{array} \right. \ .
\eea
In the case of gauge Higgs theory, our claim is that the phase of broken symmetry in the Higgs phase.

\subsection{Two Theorems}

Now let $F(U)=0$ be any physical gauge condition imposed on spacelike links, separately on each timeslice. 
We will call this an $F$-gauge.  Examples include Coulomb gauge, the Laplacian version \cite{Vink:1992ys} of Coulomb gauge,
and axial gauge.  We can prove two theorems relating $\langle \phi \rangle_F$ in physical $F$-gauges to our spin wave
order parameter $\langle \Phi \rangle$.

\begin{theorem}

In any physical $F$ gauge
\beq
          \langle \Phi \rangle \ge  |\langle \phi \rangle_F| \ .
\eeq
\end{theorem}
 
\begin{theorem}
There exists at least one physical $F$ gauge, call it $\widetilde{F}$, which saturates the bound:
\beq
          \langle \Phi \rangle =  | \langle \phi \rangle_{\widetilde{F}} | \ .
\eeq
 \end{theorem}

\noindent Custodial symmetry breaking is therefore a  necessary condition for $\langle \phi \rangle_F \ne 0$ in
any $F$-gauge, and a sufficient condition for $\langle \phi \rangle_F \ne 0$ in some $F$-gauge.  For proofs of
these statements, see \cite{Greensite:2020nhg}.

\subsection{\label{numbers} Numerical Evaluation}

\begin{figure*}[t!]
\subfigure[~]
{
\label{k1}
\includegraphics[scale=0.55]{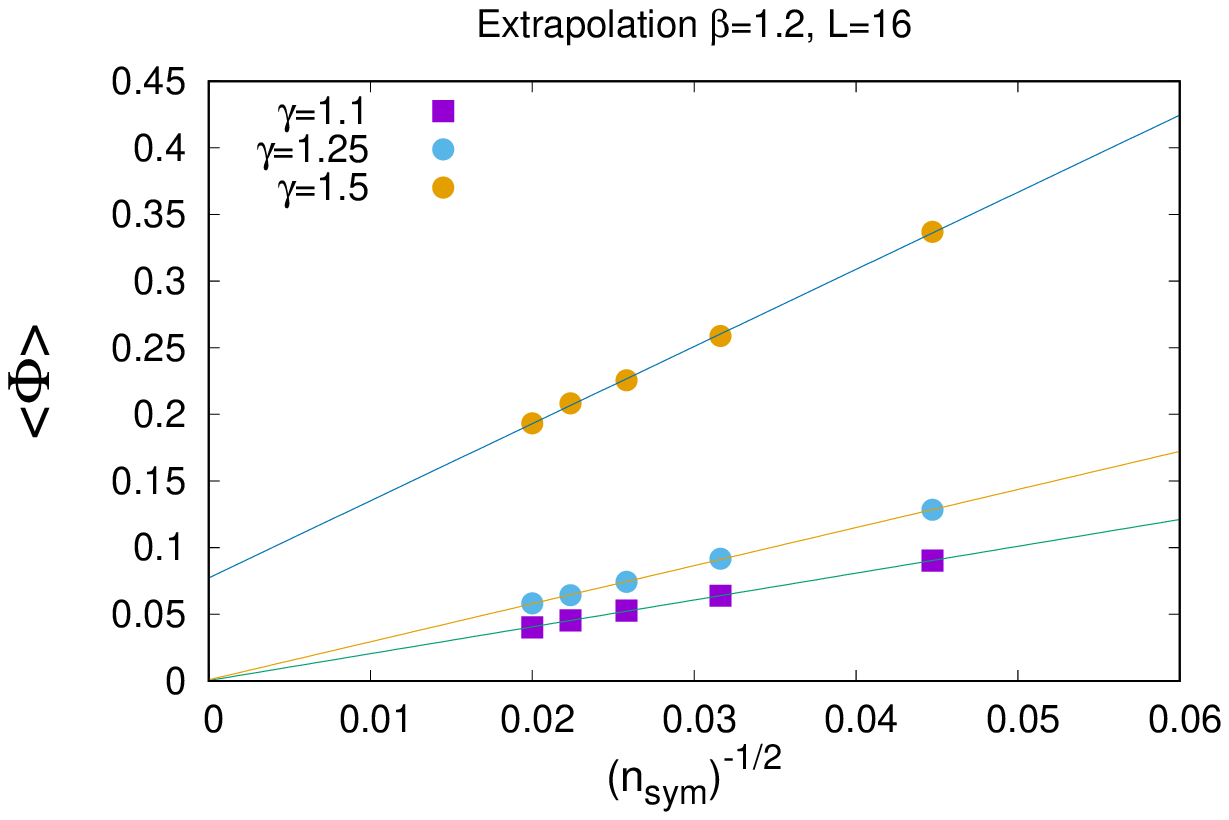}   
}
\hfill
\subfigure[~]
{   
\label{k2}
 \includegraphics[scale=0.55]{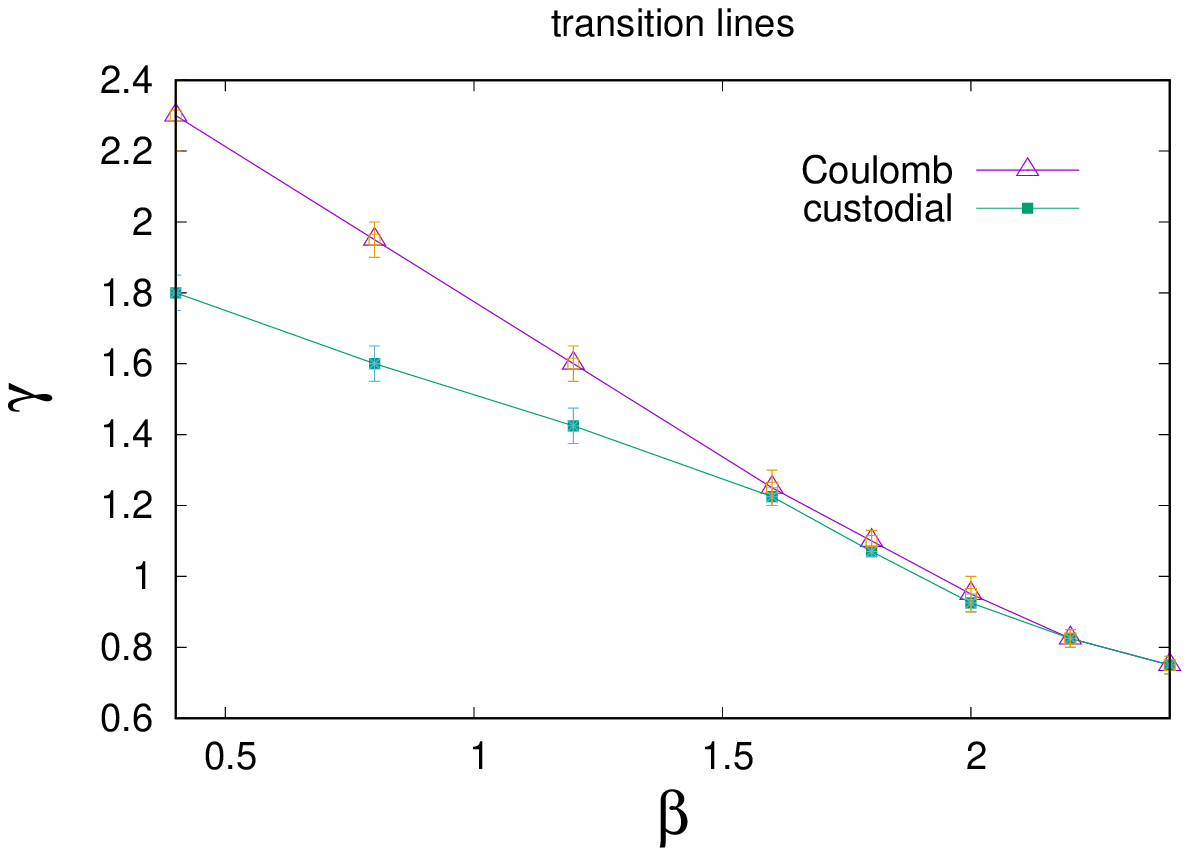}
}
\caption{(a) Extrapolation of $\langle \Phi \rangle$ to $n_{sym}\ra \infty$ above ($\g=1.5$) and below ($\gamma=1.1,1.25$) the custodial symmetry breaking transition at $\b=1.2,\g=1.4$, in SU(2) gauge Higgs theory. The lattice volume is $16^4$; error bars are smaller than the symbol sizes.  (b) The custodial symmetry breaking/spin glass transition line joins the filled squares; the Coulomb gauge transition line, joining the open triangles, lies entirely within the broken custodial symmetry phase, as it must from Theorem 1. Figures from ref.\ \cite{Greensite:2020nhg}.} 
\label{compPhi}
\end{figure*}

$\langle \Phi \rangle$ can be evaluated numerically by lattice Monte Carlo, and for practical purposes one can dispense with the complicated term in \rf{Phi} proportional to $h$.  In that case \rf{Phi} simplifies to
\bea   
       \langle \Phi[U(0)] \rangle &=& {1\over Z} \int DU_\m D\phi ~ \Phi[U(0)] e^{-S} \non \\
       \Phi[U(0)] &=& {1\over V_3} \sum_{\vx} | \overline{\phi}(\vx;U(0))| \non \\
       \overline{\phi}(\vx;U(0) )  &=&    {1\over Z_s} \int D\phi DU_0 [DU_i]_{t\ne 0} ~ \phi(\vx,t=0) e^{-S} \non \\
       Z_s &=&  \int D\phi DU_0 [DU_i]_{t\ne 0} ~ e^{-S} \ .
\label{Phi1}
\eea
Once we put the expressions in this form, we can forget about the spin glass analogy and simply regard the calculation as evaluating
the vacuum expectation of a gauge invariant operator $\Phi(U)$, which depends only on spacelike links on the $t=0$ timeslice, and which  happens to have an involved definition which itself involves a path integral. Then the order parameter can be computed with no difficulty at any $\b, \g$, via a ``Monte Carlo within a Monte Carlo'' procedure.
Figure \ref{k1} shows how this is done.    
The procedure is to update both the link and scalar field variables as usual, for e.g.\ 100 sweeps, generating configurations drawn from the usual
probability distribution $ e^{-S}/Z$.  But computation of the order parameter at each data taking ``sweep"
(better to call it an ``event") is also done by Monte Carlo simulation.   Each data taking event involves $n_{sym}$ sweeps over the lattice, where the spacelike link variables on one time slice, say $t=0$, are held fixed, while all other field variables are updated.  During these $n_{sym}$ sweeps we evaluate the average value $\overline{\phi}(\vx;U)$ at each site on the timeslice,
and from there we compute $\Phi[U]$ as in defined in \rf{Phi1}.  Averaging over many of these data taking events, with different $U_i(\vx,0)$ held fixed, gives us an estimate for $\langle \Phi \rangle$ using $n_{sym}$ sweeps in each procedure.  On general statistical grounds
\beq
           \langle \Phi \rangle_{n_{sym}} = \langle \Phi \rangle + {\text{const.} \over \sqrt{n_{sym}}} \ .
\eeq
Having computed $\langle \Phi \rangle_{n_{sym}} $ at a variety of $n_{sym}$ values, the last step is to extrapolate the data
to $n_{sym}=\infty$, as shown in Figure \ref{k1}.  At couplings inside the unbroken phase, the data extrapolates to zero.  In the broken
phase $\langle \Phi \rangle_{n_{sym}}$ extrapolates to a non-zero value.\footnote{Of course, on a finite lattice, the order parameter always vanishes for $h=0$, and what that means numerically is that without the explicit term breaking term proportional to $h$, for $n_{sym}$ large enough, eventually the data points would depart from the extrapolated line derived from the lower $n_{sym}$ values, and fall to zero.  But this departure towards zero, in the broken phase, would occur at ever larger values of $n_{sym}$, as the lattice volume is increased.}   One can estimate the point at which the transition from a zero to
non-zero value occurs, and this is the transition point.  The figure shows some sample extrapolations in SU(2) gauge Higgs theory at several $\gamma$ values, on a $16^4$ lattice, with $\beta$ is held fixed at  $\b=1.2$.  By this procedure, one arrives at a transition line (green data points) shown in Figure \ref{k2}.  The upper line is the transition line for spontaneous breaking of the ``remnant" symmetry which is left unfixed in Coulomb gauge.  This line lies above the breaking of custodial symmetry determined by the order parameter $\Phi$, in conformity with Theorem 1, i.e.\ custodial symmetry may be broken where remnant symmetry in an F-gauge (such as Coulomb) is unbroken, but not the other way around.\footnote{One should keep in mind that $\langle \phi \rangle_F$ is really the expectation value of a highly non-local quantity, namely $\langle  g_F(\vx;U) \phi(x) \rangle$, where $g_F$ is the gauge transformation to the F-gauge.}

    The gauge invariant order parameter $\Phi$ gives us an unambiguous criterion for the breaking of custodial symmetry, which includes
at a minimum the global center symmetry of the gauge group.  We have suggested that the phase of broken custodial symmetry is
the Higgs phase, but that only makes sense if there is a genuine physical distinction which corresponds to this criterion.  One might object on a priori grounds that the fact that the confinement and Higgs phases are not entirely isolated from one another by a thermodynamic transition \cite{Osterwalder:1977pc,Fradkin:1978dv,Banks:1979fi} already rules out any such essential distinction, but one
should be a little wary of this argument.  We already know of examples where there exist distinct phases of many-body systems  which are not separated by a thermodynamic transition.  One example is the roughening transition in Yang-Mills theory.  In the rough phase, the width of flux tubes grows logarithmically with quark-antiquark separation, and the static quark potential contains a $1/r$ term of stringy origin.  Neither of these features hold outside the rough phase.  Another example is the Kertesz line in the Ising model in an external field \cite{Kertesz}, which has to do with a percolation transition in the random cluster formulation of the model.  The point here is that analyticity of a local observable like the free energy does not rule out non-analytic behavior in non-local observables, and such observables can be physically important.  So our next task is to explain what, precisely, is the physical distinction between the broken and unbroken phases of custodial symmetry, and why these deserve to be called the Higgs and confinement phases respectively.

\section{Separation of charge confinement}

   We return to this question: What is the meaning of the word confinement, when there are matter fields
in the fundamental representation of the gauge group?  We know that both the Higgs and confinement phases have C confinement, since all asymptotic particle states are color neutral in both phases.  Yet there would \emph{appear} to be some qualitative differences. 
In the Higgs phase of an SU(2) gauge Higgs theory, as we know from both perturbation theory and experiment:
\begin{enumerate}
\item There are only Yukawa forces.
\item There are no linear Regge trajectories.
\item There is no flux tube formation, even as metastable states.
\end{enumerate}
and this seems to distinguish physically the Higgs from the confinement phase.  But can we make the distinction precise?

\begin{figure}[htbp]
\includegraphics[width=0.35\textwidth]{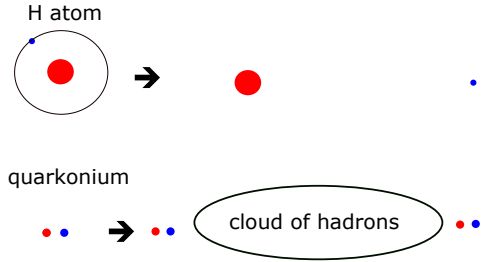} 
\caption{Unlike Hydrogen, where the ionization energy can be measured experimentally, there is no experimental procedure for
creating an ``ionized" hadron.  There are, however, physical states (see next figure) in the Hilbert space which do correspond to widely separated but interacting quarks.}
\label{Hydrogen}
\end{figure}

   Let us start with a (superficially) silly question:  what is the binding energy of the proton?  Or the J/$\psi$, or any hadron.  Obviously,
unlike the Hydrogen atom we cannot ionize a proton, or quarkonium, at least experimentally, and compare the energies of the bound and ionized states.  Instead of an isolated quark and antiquark, we get instead a bunch of color neutral hadrons (Figure \ref{Hydrogen}) with
integer electric charges.

\begin{figure}[htbp]
\includegraphics[width=0.25\textwidth]{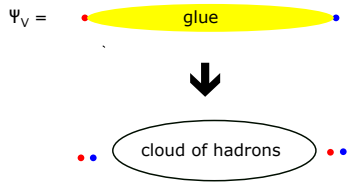} 
\caption{Decay of a state with widely separated quark-antiquark color charges and fractional electric charge into
a set of color neutral hadrons of integer electric charge.  The property of \Sc\ confinement is related to the 
energy of the color charge separated state $\Psi_V(R)$, in the limit of color charge separation $R \ra \infty$.}
\label{quarkonium}
\end{figure}

\bigskip
But there \emph{are}, nonetheless, physical states
in the Hilbert space which {\sl do} correspond to isolated (``ionized") quarks, separated by a large distance.  
Such states would be difficult to realize experimentally, but they do exist in the Hilbert space.  
For a $q\overline{q}$ system, such states have the form
\beq
     \Psi_V \equiv  \overline{q}^a(\vx) V^{ab}(\vx,\vy;A) q^b(\vy) \Psi_0 
\label{PV}
\eeq
 where  $\Psi_0$ is the ground state, $a,b$ are color indices, and $V(\vx,\vy;A)$ is a gauge bi-covariant operator which is a functional of \emph{only} the gauge field, transforming as 
\beq
V(\vx,\vy;A) \ra   g(\vx) V(\vx,\vy;A) g^{\dg}(\vy) \ .
\eeq
In QCD  there would be a fractional electric charge at $\vx$, and an opposite fractional electric charge at $\vy$, with no electric charge in between.  Of course the system would very rapidly decay into integer-charged hadrons (Figure \ref{quarkonium}).
Let $E_V(R)$ be the energy expectation value above the vacuum energy ${\cal E}_{vac}$,
\beq
      E_V(R) =   \langle \Psi_V | H | \Psi_V \rangle - {\cal E}_{vac}  \ ,
\eeq
of a state of the form \rf{PV}.

\begin{definition}
A gauge theory has the property of {\it separation-of-charge confinement} if the following
condition is satisfied:
\beq
    \lim_{R\ra \infty} E_V(R) = \infty 
\eeq
for \underline{any} choice of bi-covariant operator $V(\vx,\vy;A)$, which is a functional of only the gauge field $A$. \\
\end{definition}

It is crucial, in this definition, that $V(\vx,\vy;A)$ depends only on the gauge field, not on any matter fields, otherwise it would
be easy to constuct a $V$ operator that would
violate the \Sc\ condition, e.g.
\beq
      V^{ab}(\vx,\vy,\phi) = \phi^a(\vx) \phi^{\dg b}(\vy)   \ .
\eeq
In that case
\beq
            \Psi_V = \{ \overline{q}^a(\vx) \phi^a(\vx) \} \times \{ \phi^{\dg b}(\vy)q^b(\vy) \}\Psi_0  
\eeq
corresponds to two color singlet (static quark + Higgs) states, only weakly interacting at large separations.  Operators $V$ of this
kind, which depend on the matter fields, are excluded;  the idea is to study the energy $E_V(R)$ of physical states with large separations $R$ of static color charges unscreened by matter fields.  This also means that the lower bound on $E_V(R)$, unlike in pure gauge theories, is {\it not} the lowest energy of a state containing a static quark-antiquark pair.  It is the lowest energy of such states when color screening by matter is excluded.

   The \Sc\ (separation of charge) condition is a much stronger condition than C confinement, and in our opinion it is the natural generalization, to gauge theories with matter fields, of the linearly rising static quark potential as a confinement criterion in pure gauge theories. \Sc\ confinement presumably holds in QCD.  What about gauge Higgs theories?   
   
   First we ask whether \Sc\ confinement exists anywhere in the phase diagram except at $\gamma=0$ (pure gauge theory).
The answer is yes.   We can show  \cite{Greensite:2018mhh} that gauge-Higgs theory is  \Sc\ confining at least in the region
\beq
       \gamma \ll \beta \ll 1  ~~~\mbox{and} ~~~ \gamma \ll {1\over 10} \ .
\eeq
This is based on strong-coupling expansions and a theorem (Gershgorim) in linear algebra.
Then does \Sc\ confinement
hold  everywhere in the $\beta-\gamma$ phase diagram? 
Here the answer is no.  We can construct $V$ operators which violate the S$_\text{c}$-confinement criterion when $\g$ is large enough
\cite{Greensite:2017ajx}.
This means that there must exist a transition between S$_\text{c}$ and C confinement.  The question is whether this transition coincides
with the spontaneous breaking of custodial symmetry.

\subsection{No \Sc\ confinement in the Higgs (spin glass) phase}
 
On the lattice, $E_V(R)$ is determined from the lattice logarithmic time derivative
\bea
       \lefteqn{E_V(R)} \non \\
       &=&  - \log \left[ {\left\langle \tr\left[U_0(x,t) V(x,y,t+1) U^\dagger_0(y,t) V(y,x,t) \right] \right\rangle 
               \over \left\langle \tr\left[V(x,y,t) V(y,x,t) \right] \right\rangle} \right]  \ , \non \\
\label{EV}
\eea
where the timelike link variables arise after integrating out the static quark fields.
In the Higgs phase, according to the previous theorems,  there is always an $F$-gauge in which $\langle \phi \rangle_F \ne 0$.  This also implies $\langle U_0 \rangle \ne 0$.  Let $g_F(\vx;U)$ be the 
gauge transformation to that gauge, and choose
\beq
           V_F(\vx,\vy,t;U) = g_F^\dg(\vx,t;U) g_F(\vy,t;U)  \ .
\label{VF}  
\eeq
Then evaluating $E_V(R)$ in that $F$-gauge we have
\bea
\lim_{R\ra \infty} E_V(R) &=& - \lim_{R\ra \infty}\log \left[ {1\over N} \langle \tr [ U^\dg_0(\vx,t) U_0(\vy,t) ] \rangle_F  \right]  \non \\
              &=& -  \log \left[ {1\over N} \tr [ \langle U^\dg_0(\vx,t) \rangle_F \langle U_0(\vy,t) \rangle_F ] \right] \non \\
              &=& \text{finite} \ ,
\label{EVF}
\eea
which demonstrates the absence of \Sc\ confinement in the spin glass phase.

    Now let us define
 \bea
        |\text{charged}_{\vx \vy} \rangle &=& \q^a(\vx) V^{ab}(\vx,\vy;U) q^b(\vy) |\Psi_0 \rangle   \non \\
        |\text{neutral}_{\vx \vy} \rangle &=& ( \q^a(\vx)\phi^a(\vx) ) (\phi^{\dg b}(\vy) q^b(\vy) )  |\Psi_0 \rangle \ ,
\label{charged}
\eea
and consider the $\vy \ra \infty$ limit.  This leaves an isolated charged fermion at $\vx$ in the charged state, and
a neutral hadron at $\vx$ in the neutral state.  Let $V=V_F$ and evaluate the overlap in the Higgs phase \bigskip
\bea
       \lim_{|\vx-\vy| \ra \infty} \langle \text{neutral} | \text{charged} \rangle 
      &\propto&    \lim_{|\vx-\vy| \ra \infty} \langle  \phi^{\dg a}(\vx) \phi^a(\vy) \rangle_F \non \\
     &=& \langle \phi^{\dg a} \rangle_F \langle \phi^a \rangle_F > 0   \ .
\label{Higgs_olap}
\eea
This means there is no essential distinction between the states we have labeled ``charged" and ``neutral."  That is a consequence
of broken global $Z_N$ gauge symmetry, with the corollary that the vacuum is not an eigenstate of zero $Z_N$ charge.

    An isolated color charged particle is the source  of a long-range color electric field, and this is ruled out if the theory is massive.  In the absence of a massless sector, both the Higgs and confinement phases are C confining, and we have just established that the Higgs phase is not \Sc\ confining.  So confinement in the Higgs phase is {\it only} C confinement.

\subsection{The symmetric phase}

This time we have
\bea
     \lefteqn{\lim_{|\vx-\vy| \ra \infty} \langle \text{neutral} | \text{charged} \rangle } \non \\
      & & \qquad \qquad \propto    \lim_{|\vx-\vy| \ra \infty} \langle  \phi^{\dg a}(\vx) V^{ab}(\vx,\vy;U) \phi^b(\vy) \rangle \non \\
      & & \qquad \qquad = \lim_{|\vx-\vy| \ra \infty} \int DU \overline{\phi^a(\vx) \phi^b(\vy)}[U] V^{ab}(\vx,\vy;U) P(U)  \ , \non \\
\eea
where
\bea
 \overline{\phi^a(\vx) \phi^b(\vy)}[U]  &=& {1\over Z_{spin}(U)} \int d\phi \phi^a(\vx) \phi^b(\vy) e^{-H_{spin}/kT}  \ . \non \\
\eea
Since custodial symmetry is unbroken for gauge configurations drawn from the probability distribution $P(U)$, it follows that
for such configurations, in the symmetric phase at $h \ra 0$,
\beq
       \lim_{|\vx-\vy| \ra \infty} \overline{\phi^a(\vx) \phi^b(\vy)}[U] = 0 \ .
\eeq
Therefore
\beq
 \lim_{|\vx-\vy| \ra \infty} \langle \text{neutral} | \text{charged} \rangle = 0 \ .
\label{conf_olap}
\eeq
Note that this result holds for {\it all}  $V$ operators (all isolated charges) in the symmetric phase, independent of any gauge choice. This is a consequence of the $Z_N$ invariance of the ground state, and it means that in the symmetric phase, unlike in the broken phase, there
is a sharp distinction between charged states and color neutral states.

   Expanding on this point, consider the $\vy \ra \infty$ limit in \rf{charged}, 
\beq
    \Psi_V =   \q(\vx) V(\vx,\infty;U) \Psi_0 \ ,
\eeq
an example being $V(\vx;\infty;U) = g_F^\dg(\vx,t;U)$, where $g_F(x;U)$ is the gauge transformation to an F-gauge.
Then, although $\Psi_V$ is a physical state and therefore satisfies the Gauss Law constraint, i.e.\ is invariant under infinitesimal gauge transformations, it transforms covariantly under the global $Z_N$ subgroup of the gauge group.  In other words, it is truly a charged state, and because of its transformation properties it is orthogonal to
any color singlet, i.e.\ uncharged state.  Again we stress that this is a consequence of the $Z_N$ symmetry of the vacuum in the
symmetric state.  In the broken (Higgs, spin glass) phase the vacuum is not an eigenstate of $Z_N$, and even if an operator acting
on the vacuum has definite $Z_N$ transformation properties, the resulting physical state does not.  In fact, this is why it is possible
for $\langle \phi \rangle_F$ to have a non-zero expectation value in some F-gauge in the broken phase, despite the fact that the $\phi$ operator transforms non-trivially under the global $Z_N$ subgroup of the gauge group, and fixing to an $F$ gauge does not alter that property.

      Charged states in the unbroken phase may be of either finite of infinite energy.  If there are no charged states of finite energy above the vacuum energy, then the system is in an \Sc\ confinement phase.  If, on the other hand, there do exist charged finite energy states, orthogonal to all neutral states, then states of this kind will necessarily appear in the spectrum.  The system cannot then be in a C confining phase, where there are no charged particles in the spectrum.  Nor can it be in an \Sc\ confining phase, where isolated charges
are all states of infinite energy.  The remaining possibility is a massless phase.  So the phase of unbroken custodial symmetry
is either \Sc\ confining, or massless.  This is consistent with the fact that $\langle \phi \rangle_F=0$ in all $F$-gauges in the symmetric phase, so there exists no sensible perturbative expansion of $\phi(x)$ around a non-zero expectation value, and no Brout-Englert-Higgs mechanism in the symmetric phase, at least not one that can be seen in any physical $F$-gauge. 

    The conclusion is that the spin glass phase is a C confinement (Higgs) phase, while the phase of unbroken custodial
symmetry may be either a massless or an \Sc \ confining phase.   In the absence of a massless phase, as in SU(2) gauge Higgs theory in $D=3+1$ dimensions, the transition from the symmetric to the spin glass phase coincides with the transition from \Sc \ confinement to C confinement.

\subsection{Examples}

 It is useful to check numerically, in some examples, that
 \begin{enumerate}
 \item The charged and neutral states defined in \rf{charged} are orthogonal in the confinement phase, and have finite overlap in the Higgs phase, as a consequence of unbroken vs.\ broken $Z_N$ symmetry;
 \item The energy of any charged state diverges with charge separation in the confinement phase, regardless of $V(\vx,\vy;U)$,  but it is always possible to find operators $V(\vx,\vy;U)$ such that the resulting state has finite energy in the same limit in the Higgs phase.
\end{enumerate}
This will be illustrated in SU(3) gauge Higgs theory.  But before proceeding to those results, let us look at the simplest possible example
of a charged state, namely pure QED with a static charge (and no dynamical charges) at the point $\vx$.  The lowest energy state
of this kind was written down long ago by Dirac \cite{Dirac:1955uv}:
\beq
            |\Psi_\vx \rangle = \overline{\psi}^\dg(\vx) \rho_C(\vx;A) |\Psi_0 \rangle \ ,
\eeq
where
\bea
            \rho_C(\vx;A) &=& \exp\left[-i {e\over 4\pi} \int d^3z ~ A_i(\vz) {\pa \over \pa z_i}  {1\over |\vx-\vz|}  \right]  \ .
\eea
It is easy to check that $ |\Psi_\vx \rangle$ satisfies the Gauss Law.  However, let $g(x) = e^{i\theta(x)}$
be an arbitrary U(1) gauge transformation, and we separate out the zero mode $\theta(x) = \theta_0 + \tilde{\theta}(x)$.
Then
\beq
          \psi(\vx) \ra e^{i\theta(\vx)}\psi(\vx) ~~~\mbox{but}~~~ \rho_C(\vx;A) \ra e^{i\tilde{\theta}(x)} \rho_C(\vx;A) \ ,
 \eeq
and therefore
\beq
|\Psi_\vx \rangle \ra e^{-i\theta_0} |\Psi_\vx \rangle \ .
\eeq 
So $|\Psi_\vx \rangle$ transforms covariantly under the global center subgroup (which is U(1)) of the U(1) gauge group; this is the
hallmark of a charged state.  The vacuum state is invariant under such transformations, simply because it depends only on the gauge field, which is itself invariant
under those global transformations. This means that $ |\Psi_\vx \rangle$ is a charged state, and of course it is finite energy (apart from
the usual UV divergence which is regulated on the lattice).  

The operator  $\rho_C(\vx;A)$ is an example of what we have elsewhere \cite{Greensite:2017ajx} called a ``pseudomatter"  field; this is a field which transforms like a matter field in the fundamental representation, except that it is invariant under transformations in the global center subgroup of the gauge group.  Any gauge transformation $g_F(x;U)$ to an F-gauge can be decomposed \\ into $N$ pseudomatter fields $\{\rho_n\}$, and vice-versa:
\beq
           \rho^a_n(\vx;A) = g_F^{\dg an}(\vx;A) \ .
\eeq
In particular, the operator $\rho_C^*(\vx;A)$ defined earlier is precisely the gauge transformation
to Coulomb gauge in an abelian theory.  This operator dresses a static charge with a surrounding Coulomb field.  Another example,
in an SU(N) lattice gauge theory, is any eigenstate $\xi_n(\vx;U)$ of the covariant Laplacian operator
\beq
-D^2 \xi_n = \kappa_n \xi_n \ ,
\eeq 
where
\bea
  (-D^2)^{ab}_{\vx \vy} =  \sum_{k=1}^3 \left[2 \delta^{ab} \delta_{\vx \vy} - U_k^{ab}(\vx) \delta_{\vy,\vx+\hat{k}} 
       - U_k^{\dg ab}(\vx-\hat{k}) \delta_{\vy,\vx-\hat{k}}  \right]    \ . \non \\
\eea
These constructions underlie the Laplacian gauge introduced by Vink and Weise \cite{Vink:1992ys}, which is free of Gribov copies.

\begin{figure*}[t!]
\subfigure[~confinement phase] 
{   
 \label{EV05}
 \includegraphics[width=8cm]{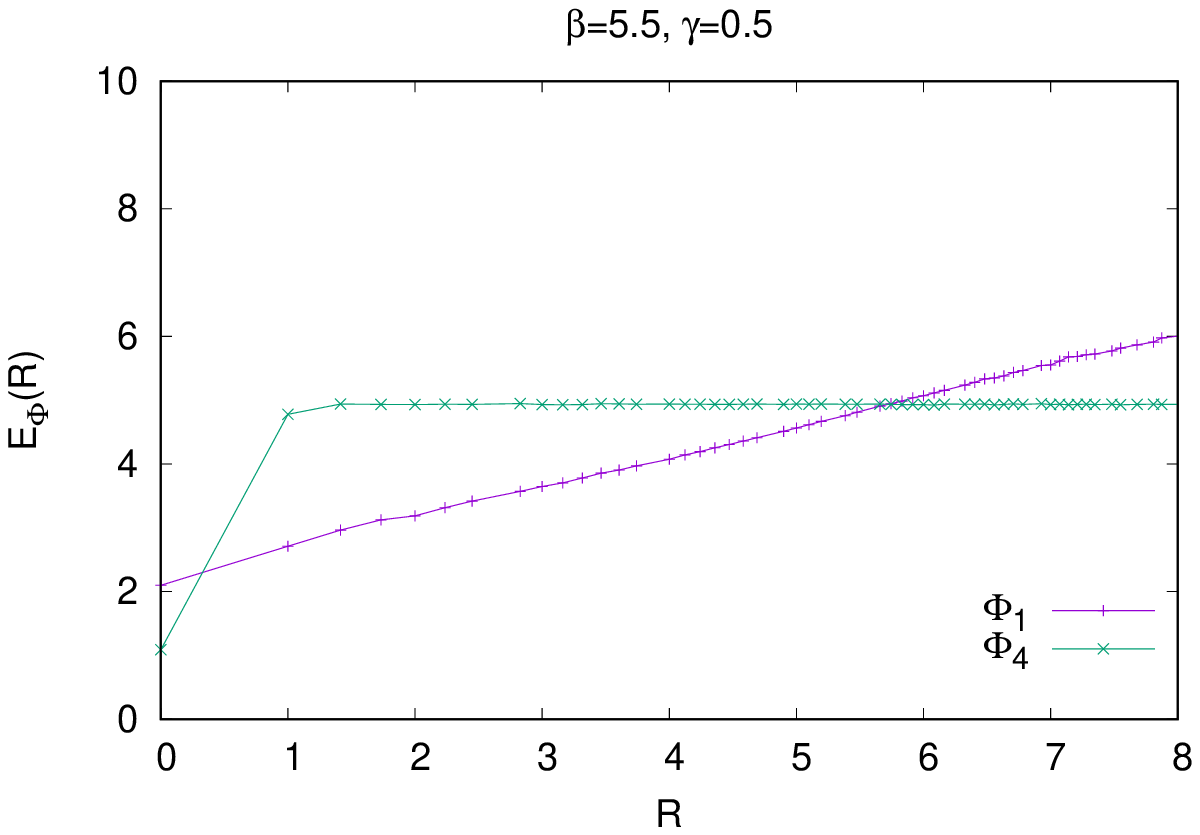}
}
\subfigure[~Higgs phase]  
{ 
 \label{EV35}
 \hspace{-20pt} \includegraphics[width=8cm]{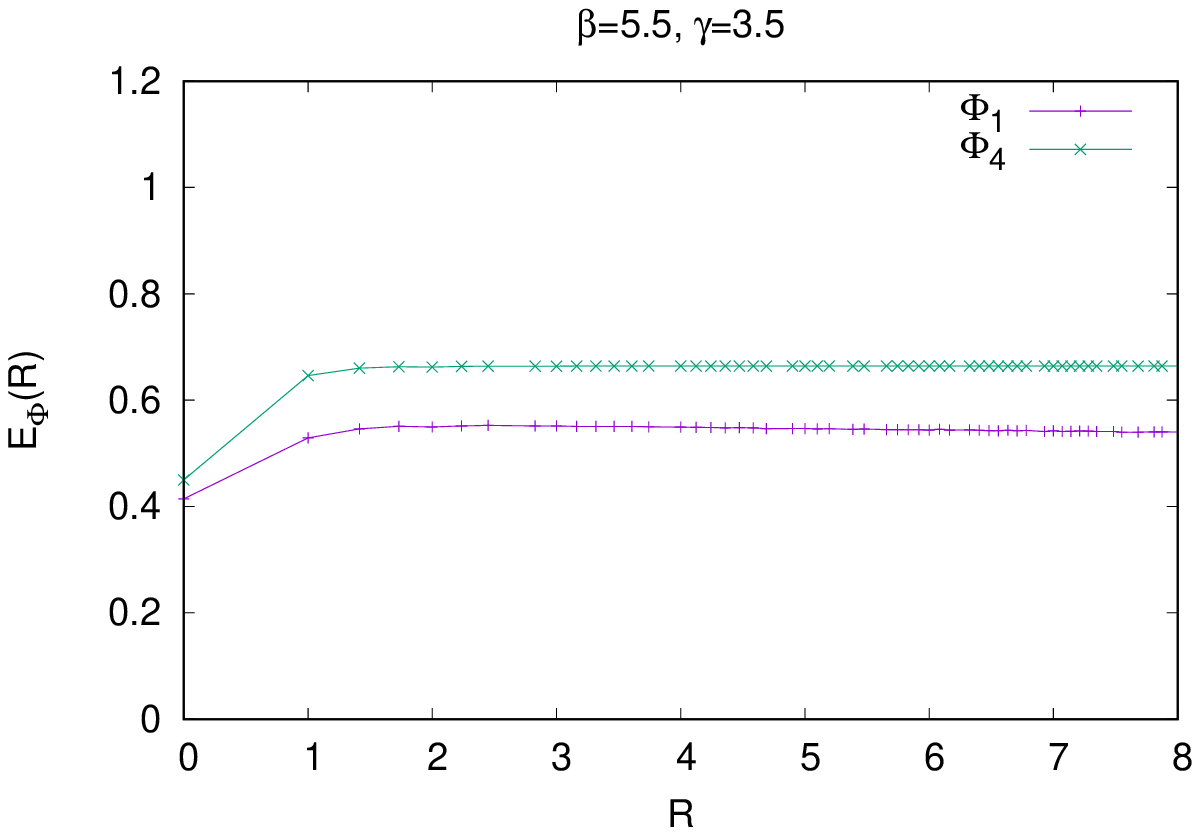}
}
\subfigure[~confinement phase] 
{   
 \label{olap1}
 \includegraphics[width=8cm]{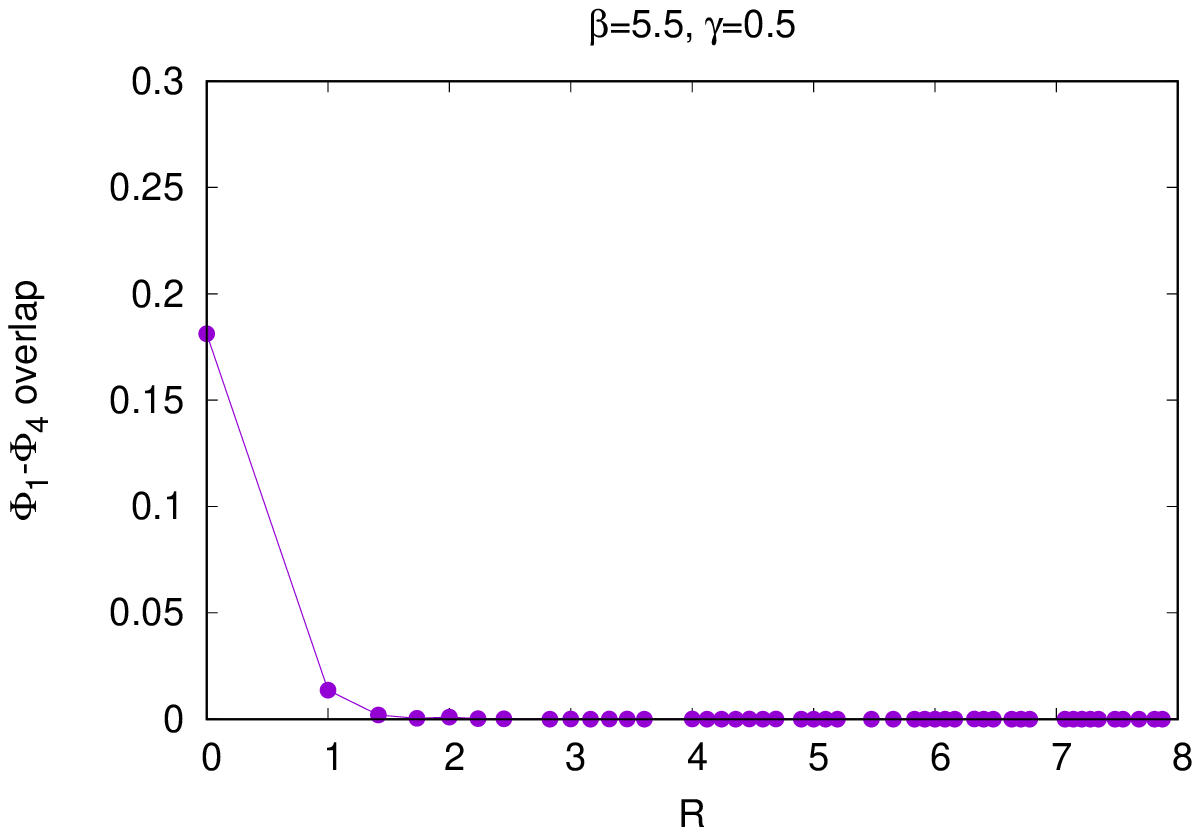}
}
\subfigure[~Higgs phase] 
{   
 \label{olap2}
 \includegraphics[width=8cm]{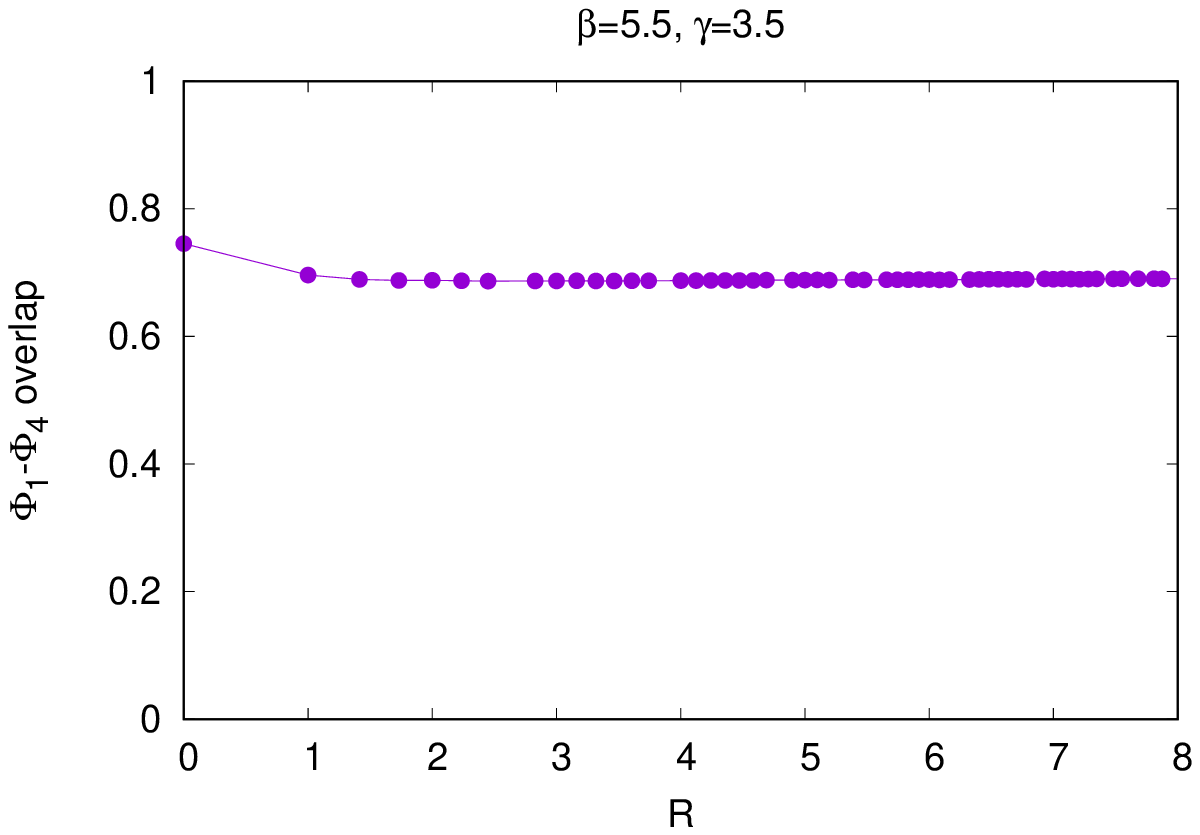}
}
\caption{Contrasting properties of charged ($\Phi_1$) and neutral ($\Phi_4$) fermion-antifermion states in the confinement and Higgs phases of an SU(3) gauge Higgs theory. (a) Energy expectation value $E^\Phi(R)$ vs.\ separation $R$ of the $\Phi_1$ and $\Phi_4$ states in the confined phase, $\b=5.5, \g=0.5$. (b) Same as subfigure (a), but in the Higgs phase at $\b=5.5, \g=3.5$.  (c) Overlap vs.\ $R$ of normalized charge ($\Phi_1$) and  neutral ($\Phi_4$) states in the confined phase, at $\b=5.5, \g=0.5$. (d) Same as subfigure (c), but in the Higgs phase at $\b=5.5, \g=3.5$. Figures from ref.\ \cite{Greensite:2020lmh}.}
\label{Psidata}
\end{figure*}

    Now let us consider an SU(3) lattice gauge Higgs theory, with action
\bea
        S &=& - {\beta \over 3} \sum_{plaq} \mbox{ReTr}[U_\m(x)U_\n(x+\hat{\m})U_\m^\dg(x+\hat{\n}) U^\dg_\n(x)]  \non \\
      & & \qquad - \gamma \sum_{x,\m} \mbox{Re}[\phi^\dg(x) U_\m(x) \phi(x+\widehat{\m})] \ , \non \\ 
\eea
and we impose for simplicity a unimodular constraint $|\phi|=1$ on the Higgs field.  
Let us define
\bea
          |\text{charged}_{\vx \vy}\rangle &=& \q^a(\vx) V_A^{ab}(\vx,\vy;U) q^b(\vy) |\Psi_0 \rangle  \non \\          
          |\text{neutral}_{\vx \vy}\rangle  &=&   \q^a(\vx)  V_B^{ab}(\vx,\vy;\phi)  q^b(\vy)  | \Psi_0 \rangle  \ ,
\eea
where
\bea
         V_A^{ab}(\vx,\vy;U) &=& \xi_1^a(\vx;U) \xi_1^{\dg b}(\vy;U) \non \\
         V_B^{ab}(\vx,\vy;\phi) &=& \phi^a(\vx) \phi^{\dg b}(\vy) \ , 
\label{VAB}
\eea         
and where the $\xi_1(\vx;U)$ is a pseudomatter operator corresponding to the eigenstate of the lattice Laplacian operator with the smallest
eigenvalue.  The energy expectation values of the charged and neutral states are then given by eq.\ \rf{EV}, using the operators in
\rf{VAB} above, and these we have computed numerically in \cite{Greensite:2020lmh}.   The results obtained in the confinement phase 
at $\beta=5.5, \gamma=0.5$,
and in the Higgs phase at $\beta=5.5, \gamma=3.5$, are shown in Figure \ref{Psidata}.  Note that in these plots, taken from \cite{Greensite:2020lmh},
the charged and neutral states are labeled
\beq
 |\text{charged}_{\vx \vy}\rangle \ra |\Phi_1(R) \rangle ~~~,~~~~  |\text{neutral}_{\vx \vy}\rangle \ra |\Phi_4(R) \rangle \ ,
 \eeq
 for reasons given in that reference.

   These plots illustrate the statements made about the distinction between different types of confinement in the Higgs and 
confinement phases.  The confined phase is \Sc\ confining.  Hence the energy of any charged state, created with \emph{any} 
$V(\vx,\vy,U)$ operator, will diverge at quark separation $R\ra \infty$.  The energy of a neutral state (in this case the charge of the quarks are 
neutralized by the Higgs fields), will be finite in the same limit. This is all clearly seen in Figure \ref{EV05}.  The overlap of charged and
neutral states vanishes in the confinement phase (a prediction of eq.\ \rf{conf_olap}), as we see in Figure \ref{olap1}.  Things are different in the Higgs
phase.  In this case the energies of both the charged and neutral states go to a finite constant as $R\ra \infty$ as seen in
Figure \ref{EV35}, meaning that the Higgs phase is {\it not} a phase of \Sc\ confinement, and in Figure \ref{olap2} we see that the overlap 
between the two states is substantial (this time a prediction of eq.\ \rf{Higgs_olap}) in the same limit, due to the fact that the the global center subgroup of the SU(3) gauge group is spontaneously broken, and there is no longer any essential distinction between charged and neutral states, so long as those states satisfy the required Gauss Law constraint.
 
\subsection{QCD as a symmetric phase}

    Since our discussion has focused on gauge Higgs theory, it is natural to ask how it applies to
a theory like QCD, where the matter in the fundamental representation of the gauge group is fermionic, and we would expect
that the global center subgroup of the gauge group is unbroken.  Although this question
has not yet been addressed numerically, we believe it might be addressed in the following way:  After integrating out the
fermionic degrees of freedom, one ends up with a fermionic determinant, which in turn can be expressed in terms of 
pseudofermion fields, which are in fact scalar fields in the fundamental representation.  This is how QCD can be simulated
numerically, via the hybrid Monte Carlo algorithm.  The pseudofermion action is, of course, non-local, but apart from computation time
we see no difficulty in principle in applying the numerical approach outlined in section \ref{numbers} to QCD, where we would expect 
$\langle \Phi \rangle = 0$ in the appropriate limit.   

\section{Conclusions}

  In this article we have argued that, contrary to the general consensus, there really is an essential distinction to be made between
the confinement and Higgs phases of a gauge Higgs theory, and that these phases are distinguished by confinement type, and by the broken or unbroken realization of a global center subgroup of the gauge group.  In this respect, which is not in contradiction to
Elitzur's theorem, the Higgs phase really is a phase of spontaneously broken gauge symmetry, and this breaking is detected by a gauge invariant order parameter closely analogous to the Edwards-Anderson order parameter for a spin glass.  In fact, we believe it is sensible to view the Higgs phase as a kind of spin glass phase of a gauge Higgs theory.

   The general consensus is also that the word ``confinement," in a gauge theory with matter in the fundamental representation, can only mean that the asymptotic particle spectrum is color neutral, which is a property we call C (color) confinement.  This property holds in both the Higgs and confinement phases, but not, of course, in a massless phase.  However, we have shown that there is a stronger variety of confinement which exists in the confinement phase,  and which distinguishes that phase physically from the Higgs phase.  This is the property of \Sc\ (separation of charge) confinement, associated with the formation of metastable color electric flux tubes, and it is
the natural extension of confinement criteria in a pure gauge theory to theories with matter fields.  As outlined here,
and in more detail in ref.\ \cite{Greensite:2020nhg}, the transition line between \Sc\ and C confinement coincides with the transition
from the unbroken to the broken phase of the global center subgroup of local gauge symmetry.

\acknowledgements{This research is supported by the U.S.\ Department of Energy under Grant No.\ DE-SC0013682.}

\bibliography{sym3}

\end{document}